\documentclass[conference]{IEEEtran}
\usepackage{amsmath,amsfonts}
\usepackage{algorithm}
\usepackage{algpseudocode}
\usepackage{array}
\usepackage[caption=false,font=normalsize,labelfont=sf,textfont=sf]{subfig}
\usepackage{textcomp}
\usepackage{stfloats}
\usepackage{url}
\usepackage{bm}
\usepackage{verbatim}
\usepackage{graphicx}
\usepackage{siunitx}
\def\BibTeX{{\rm B\kern-.05em{\sc i\kern-.025em b}\kern-.08em
    T\kern-.1667em\lower.7ex\hbox{E}\kern-.125emX}}
\usepackage{balance}
\usepackage{comment}

\newcommand{\N}{\mathcal{N}}
\newcommand{\Lines}{\mathcal{L}}
\newcommand{\T}{\mathcal{T}}
\newcommand{\B}{\mathcal{B}}
\newcommand{\storage}{\mathcal{J}}

\newcommand{\pcharge}{P^{\texttt{ch}}_{j,t}}
\newcommand{\pdischarge}{P^{\texttt{dis}}_{j,t}}

\newcommand{\pred}{P^{\texttt{DR}}_{b,t}}
\newcommand{\demand}{d_{b,t}}

\newcommand{\pricebuy}{\lambda^{\texttt{buy}}_t}
\newcommand{\priceflex}{\lambda^{\texttt{flex}}_t}
\newcommand{\pricered}{\lambda^{\texttt{DR}}_t}

\begin{document}
\title{Safe Bottom-Up Flexibility Provision from Distributed Energy Resources}

\author{
Costas Mylonas\textsuperscript{a, c},
Emmanouel Varvarigos\textsuperscript{a},
Georgios Tsaousoglou\textsuperscript{b}\\[1ex]
\textsuperscript{a}Department of Electrical and Computer Engineering, National Technical University of Athens, Athens, Greece\\
\textsuperscript{b}Department of Applied Mathematics and Computer Science, Technical University of Denmark, Lyngby, Denmark\\
\textsuperscript{c}Energy Digitalization Group, UBITECH, Athens, Greece
}



\maketitle

\begin{abstract}
Modern renewables-based power systems need to tap on the flexibility of Distributed Energy Resources (DERs) connected to distribution networks. It is important, however, that DER owners/users remain in control of their assets, decisions, and objectives. At the same time, the dynamic landscape of DER-penetrated distribution networks calls for agile, data-driven flexibility management frameworks. In the face of these developments, the Multi-Agent Reinforcement Learning (MARL) paradigm is gaining significant attention, as a distributed and data-driven decision-making policy. This paper addresses the need for bottom-up DER management decisions to account for the distribution network's safety-related constraints. While the related literature on safe MARL typically assumes that network characteristics are available and incorporated into the policy's safety layer, which implies active DSO engagement, this paper ensures that self-organized DER communities are enabled to provide distribution-network-safe flexibility services without relying on the aspirational and problematic requirement of bringing the DSO in the decision-making loop.
\end{abstract}


\section{Introduction}

\subsection{Motivation}
Distributed flexibility is envisioned as a key driver towards reliable renewables-based power systems.
Demand-response frameworks draw on the aggregated flexibility of multiple Distributed Energy Resources (DERs) to provide balancing/regulation services, while simultaneously respecting the DERs' operational constraints and their owners' preferences. The effective management of a large number of small-scale DERs, owned and operated by private prosumers, presents scalability and privacy challenges.

Naturally, under high renewable energy penetration, the decisions over activating the finite DER flexibility must be made stochastically and adaptively.
Namely, flexibility activation decisions should account for future-related effects, such as the depletion of stored energy in batteries and the exhaustion of buildings' indoor temperature tolerance, in par with the predicted need and value of further flexibility activation in subsequent hours.
This constitutes the DER dispatch a problem of \textit{sequential decision-making}, where the solution concept takes the form of a decision-making policy, i.e. a method for making a decision at each stage, once faced with the current system state.

An additional requirement is that the DERs' dispatch decisions need to be coordinated to ensure efficiency and satisfaction of coupled constraints; namely, preventing reverse-peak effects and violations of distribution network safety constraints. 
At the same time though, each entity (e.g. building, storage facility, etc) should remain in control of its own assets and entitled to pursue its own preferences and objectives, which challenges traditional top-down operational frameworks.

While DER owners are not inherently responsible for voltage control, future regulatory frameworks and emerging ancillary service markets may reward prosumers for grid-friendly behavior \cite{ceer2020ceer}. Moreover, ensuring that DER flexibility provision (to the wholesale market) does not create operational issues —such as voltage violations or reverse flows— in the underlying low-voltage network, adds further motivation for localized voltage-aware decision-making, even without direct DSO coordination.

Thereupon, bottom-up, \textit{distributed} coordination schemes are motivated, to achieve coordinated outcomes without requiring that DER owners lose control over their assets to a central entity, e.g., an Aggregator or a Distribution System Operator (DSO).
%
In summary of the above, in this paper we are dealing with the need to design a safe, jointly efficient, and bottom-up sequential-decision-making policy, for a cluster of independent DERs connected to a common power distribution network.

\subsection{Related Work}

The need for distributed decisions that respect distribution network constraints, points to studies proposing distributed algorithms for solving the AC optimal power flow (OPF) problem. A comprehensive review of relevant methodologies is presented in \cite{molzahn2017survey}. Although \cite{molzahn2017survey} focuses on deterministic instances of the OPF, those have been extended to sequential decision-making frameworks: A typical approach to designing a policy follows the so-called ``predict-then-optimize'' procedure, where some forecasting model is used to produce (probabilistic or point-) forecasts of the uncertainties, which is then fed into a (stochastic or deterministic) optimization problem, with the process repeating in a rolling-horizon fashion (refer e.g. to \cite{tsaousoglou2023integrating} and references therein).

As the landscape of distribution systems is rapidly evolving, the above model-based approaches require repeated efforts in model reconfiguration and staying up to date. 
This, along with the difficulty to design nuanced models for every individual use case, has motivated a major pivot towards self-adapting data-driven approaches, the predominant representative of which being Multi-Agent Reinforcement Learning (MARL).

The authors of \cite{zhang2022multi} propose a MARL approach for coordinating the operation of multiple buildings in a grid-aware manner. The proposed approach enables each building to independently learn its optimal control policy while considering the overall system state. 
In \cite{charbonnier2022scalable}, a scalable MARL approach for distributed control of residential energy flexibility is presented. The authors highlight the ability of the proposed approach to handle large-scale problems and its robustness to changes in the environment. Authors of \cite{ye2022multi} present a MARL approach for coordinated energy trading and flexibility services provision in local electricity markets. 
These studies have demonstrated the potential of MARL to coordinate the operation of multiple agents in a manner that optimizes energy trading, reduces total energy cost, and improves the reliability of the power grid. However, a common limitation of these approaches is the potential for constraint violations.

Acknowledging the need to ensure constraint satisfaction, inherent in safety-critical applications, recent AI literature has made significant strides towards ensuring safety within MARL. Notable examples include the shielding technique of \cite{elsayed2021safe}, the incorporation of safety directly into the policy search, such as 
\cite{gu2021multi} and \cite{lu2021decentralized}
and safe MARL algorithms (cf 
\cite{sheebaelhamd2021safe}) for continuous action spaces, using a safety layer.
%
Inspired by these techniques, researchers have recently applied safe data-driven methods to power systems. Namely,
\cite{wang2023secure} proposes a safe reinforcement learning approach for secure energy management of multi-energy microgrids,
while \cite{gao2022model} presents a model-augmented safe reinforcement learning approach for Volt-VAR control. 
These approaches, however, adopt a single-agent perspective, resulting in centralized decision-making where a central entity governs the entire network to optimize a global objective.

Recent advancements in safe MARL algorithms have significantly contributed to the development of scalable, distributed, and secure energy management systems within distribution networks. In \cite{zhang2024networked}, the authors introduce a networked safe MARL framework aimed at low-carbon demand management. 
Authors of \cite{chen2022physics} propose a physics-shielded MARL strategy for active voltage control.
The core of their methodology involves embedding physical constraints directly into the learning process, effectively creating a ``shield'' that guides the agents to operate within safe states.
\cite{guo2023safe} 
proposes a safe MARL algorithm for real-time Volt-Var control in distribution grids, aiming to minimize losses while maintaining nodal voltages within a safe range. 
In \cite{zhang2020multi}, the authors 
present a safe MARL policy that optimizes energy distribution and consumption across microgrids, 
enabling decentralized decision-making while incorporating safety constraints into the learning process.

While these studies represent notable advancements in multi-agent deep reinforcement learning for energy systems, they exhibit some drawbacks. Specifically, the methodologies employed by  \cite{zhang2024networked} and \cite{zhang2020multi} rely on satisfying operational constraints in expectation rather than ensuring their consistent adherence, which cannot guarantee constraint satisfaction at all times. 
Furthermore, \cite{guo2023safe} assumes prior knowledge of network parameters for the implementation of the safety layer optimization. This assumption could limit the applicability of the model in real-world scenarios where such parameters may not be readily available or could change dynamically, thereby affecting the model's robustness and adaptability. Additionally, \cite{chen2022physics} incorporates a penalty term from the safety shield directly into the reward function, along with a coefficient requiring fine-tuning. This approach introduces complexity into the reward structure, necessitating meticulous calibration to balance safety and performance objectives effectively.


\subsection{Research Gap and Contributions}

In addition to the specific challenges of each individual approach discussed, an overarching shortcoming of the presented literature is the assumption that distribution network constraints can be evaluated during training or execution. The practical implication is that the DSO is provisioned to either give away the distribution network topology and characteristics to the private asset owners, or be actively engaged into the decision-making loop.

In addressing the identified gaps, this paper presents a safe MARL framework for safe and bottom-up flexibility provision of DERs (specifically storage systems and flexible prosumers/buildings) in distribution networks.
The proposed framework features:
\begin{itemize}
    \item decentralized decision-making, with each prosumer remaining in control of its own assets and objectives;
    \item accounting for intertemporal constraints such as limited cumulative building flexibility and end-of-horizon storage level constraints;
    \item ensuring the satisfaction of voltage limits without assuming access to the network topology and characteristics.
\end{itemize}
The latter feature is achieved by training a model-free regressor, that is able to predict voltages as a function of flexibility actions based purely on data, and incorporating it into the safety layer of the MARL policy.
The practical implication is that self-organized DER communities are enabled to provide distribution-network-safe flexibility services without relying on the aspirational and problematic requirement of active, in-the-loop, DSO engagement.

\section{System Model}

We consider a set of flexible DERs, connected to a power distribution network, within a time horizon $\T$. These DERs include a set $\storage$ of Energy Storage Systems (ESS) and a set $\B$ of flexible buildings that provide load reduction services within the specifications of a Demand Response (DR) contract.

\subsection{DER Models}

\noindent \textbf{An ESS} $j \in\storage$ is characterized by the stored energy dynamics equation:
\begin{equation} \label{storage_dynamics}
e_{j,t+1} = e_{j,t} + \Delta t \cdot \big(\eta^{\texttt{ch}}_j \cdot \pcharge - \frac{1}{\eta^{\texttt{dis}}_j} \cdot \pdischarge\big),~ \forall j \in\storage, t \in \T,
\end{equation}
\noindent where $\pcharge$ and $\pdischarge$ are the charging and discharging power of storage $j$ at time $t$, $e_{j,t}$ is the energy level, $\eta^{\texttt{ch}}_j$ and $\eta^{\texttt{dis}}_j$ are the charging and discharging efficiencies and $\Delta t$ is the time step duration.
The charging and discharging powers as well as the stored energy are subject to operational limits:

\begin{align}
\label{eq:ess_charging_bounds}
&0\leq \pcharge \leq \overline{\mathrm{P}}^{\texttt{ch}}_j,
\quad &\forall j \in\storage, t \in \T, \\
\label{eq:ess_discharging_bounds}
&0 \leq \pdischarge \leq \overline{\mathrm{P}}^{\texttt{dis}}_j,
\quad &\forall j \in\storage, t \in \T,\\
\label{eq:ess_energy_bounds}
&\underline{\mathrm{E}}_j \leq e_{j,t} \leq \overline{\mathrm{E}}_j, \quad &\forall j \in\storage, t \in \T.
\end{align}
Finally, to avoid myopic behavior, the ESS must retain at least half of its maximum capacity by the end of the optimization horizon:
\begin{equation} \label{eq:storage_horizon_end}
e_{j, |\T|} \geq 0.5 \cdot \overline{\mathrm{E}}_j, \quad \forall j \in\storage.
\end{equation}

\noindent \textbf{A building} $b \in \B$ is characterized by a net energy demand $\demand$ (energy consumption minus possible generation from on-site renewables). This demand can be reduced upon request, by an amount $\pred$, which is subject to a bound:
%
\begin{equation}
\label{eq:building_bounds}
0 \leq \pred \leq \overline{\mathrm{P}}^{\texttt{DR}}_b, \quad \forall b \in \B, t \in \T,
\end{equation}
\noindent where $\overline{\mathrm{P}}^{\texttt{DR}}_b < \demand$ depends on the building's operational flexibility and the DR contract. 
The contract also stipulates a limit on the cumulative reduction over the entire scheduling horizon:
\begin{equation}
\label{eq:total_reduction_constraint}
\sum_{t \in \T} \pred \leq \mathrm{H}_b \cdot |\T| \cdot \overline{\mathrm{P}}^{\texttt{DR}}_b, 
\quad \forall b \in \B, t \in \T,
\end{equation}

\noindent where \(\mathrm{H}_b \in [0,1]\) is a parameter indicating the maximum fraction of building \(b\)'s total theoretical reduction capacity that can be utilized over the horizon $\T$.

\subsection{Distribution Network Model}

We now turn to the operational constraints of the power distribution network that hosts the DERs.
The network comprises a set $\N$ of nodes and a set $\Lines$ of interconnecting lines. We assume one ESS and one building at each node, without loss of generality, to avoid carrying additional notation (i.e. separate sets and sums for the DERs of each node) in the formulas that follow. Accordingly, we define the active power balance equation for node $n$ as
\begin{equation} \label{power_balance}
\demand - \pred + \pcharge - \pdischarge = \sum_{l \in \Lines_n}\!P_{l,t}, \quad \forall n \in \N, t \in \T,
\end{equation}

\noindent where $\Lines_n$ is the set of lines that node $n$ is connected to, and $P_{l,t}$ is the power flow (positive or negative depending on the direction) for line $l$.
The reactive power balance equation reads
\begin{equation}
\sum_{l \in \Lines_n}\!Q_{l,t} = q_{n,t}, \quad \forall n \in \N, t \in \T,
\end{equation}

\noindent where \(q_{n,t}\) includes the total reactive power absorption at bus \(n\) at time \(t\).
We define the variables \(v_{n,t} = \bigl(V_{n,t}\bigr)^2\) to capture the squared voltage at bus $n \in \N$ and \(\ell_{l,t} = \bigl(I_{l,t}\bigr)^2\) to capture the squared current on line \(l\in L\). The voltage magnitude limits are imposed in terms of squared voltages:
\begin{equation}
\label{eq:volt_limits_socp}
\underline{\mathrm{V}} \leq v_{n,t} \leq \overline{\mathrm{V}}, 
\quad \forall n \in N, t \in \T.
\end{equation}
%

%


\noindent For two directly connected buses $n, m$, the voltage drop from bus \(n\) to bus \(m\) is given by
\begin{equation}
\begin{aligned}
v_{m,t} \;=\; v_{n,t} -\; 2\,\bigl(\mathrm{R}_l\,P_{l,t}\;+\;\mathrm{X}_l\,Q_{l,t}\bigr) + \;\bigl(\mathrm{R}_l^2 + \mathrm{X}_l^2\bigr)\,\ell_{l,t}, \\
\quad \forall n, m: l \in \Lines_n \cap \Lines_m, t\in \T,
\end{aligned}
\end{equation}

\noindent where $\mathrm{R}_l$ and \(\mathrm{X}_l\) are the resistance and reactance of line \(l\), while current magnitudes are given by

\begin{equation} \label{currents}
P_{l,t}^2 + Q_{l,t}^2 \;\;=\;\;\ell_{l,t}\,v_{n,t}, \quad \forall\,l\in \Lines_n, t\in \T.
\end{equation}





\subsection{Problem Formulation}

The DERs participate (through an aggregator/community) in the real-time balancing market by providing up-regulation services, where up-regulation (elicited by discharging the ESS and reducing the buildings' energy consumption) is compensated at a per-unit price $\priceflex$. Information on this price becomes available only in real time.
Additionally, the cost of energy consumption is based on a standard tariff $\pricebuy$.
Finally, reducing a building's consumption comes at a per-unit cost of $\pricered$, per the 
building's DR contract.

The objective is to maximize the total net benefit from flexibility provision (i.e. revenue from providing regulation services minus cost from consumption and from procuring flexibility from buildings and ESSs) without violating the distribution network safety constraints. The latter requirement is particularly challenging, given that the DERs are not aware of the distribution network's topology and line characteristics, but only have access to local voltage measurements. We define the described problem as a Constrained Markov Decision Process $\mathcal{M}$:

\begin{itemize}

\item State Space \(\mathcal{S}\): We define the global state at time \( t \) over the state variables $(e_{j,t})_{j \in \storage}, (\demand)_{b \in \B}, (q_{n,t})_{n \in \N},$ and $(\lambda^{\texttt{flex}}_{\tau})_{\tau \in \mathcal{H}_t}$ which encodes the historical regulation prices for some look-back horizon $\mathcal{H}_t$ up to (and including) $t$.





\item Action Space $\mathcal{A}$: The action variables include the charge, discharge and DR decisions $\pcharge, \pdischarge, \pred$ for all buildings and ESSs.

\item Reward Function: The reward function is defined as
\begin{multline} \label{eq:reward}
    r_t = \priceflex\Big(\sum_{b \in \B}\pred + \sum_{j \in \storage}\pdischarge\Big) - \\
    \pricebuy\Big(\sum_{j \in \storage}\pcharge + \sum_{b \in \B} (\demand - \pred)\Big) - \\
    \pricered \sum_{b \in \B}\pred. 
\end{multline}

\item Transition Function: The exogenous state variables $\demand, q_{n,t}, \priceflex$ evolve based on an unknown stochastic process (instantiated by historical data). The ESS stored energy evolves deterministically per Eq. \eqref{storage_dynamics}.


\item Constraints: The MDP's constraints refer to operational bounds \eqref{eq:ess_charging_bounds}-\eqref{eq:total_reduction_constraint}, and the voltage limits \eqref{eq:volt_limits_socp}, where the voltages result through the power flow equations \eqref{power_balance}-\eqref{currents}. 


\end{itemize}

The C-MDP $\mathcal{M}$ is particularly challenging due to the continuous and high dimensional state-action space, the intertemporal nature of constraints \eqref{eq:storage_horizon_end}, \eqref{eq:total_reduction_constraint}, and especially the voltage limits' constraint, given that the DERs are ignorant of the distribution network topology and line characteristics.
Finally, the requirement that each node remains in control of its own assets mandates a distributed solution.

\section{Proposed Solution}

\subsection{Multi-Agent Deep Reinforcement Learning}
Towards handling the distributed, sequential decision-making problem presented, which bears a continuous and high-dimensional state-action space, we adopt a 
Safe Multi-Agent Deep Reinforcement Learning (MADRL) approach, where an agent is associated with the decisions of the assets under a particular distribution network node. 
An agent's observations $o_{i,t}$ follow directly from the state space of $\mathcal{M}$ (by constricting the agent's scope to only the assets of a single node).
To facilitate the algorithm's performance, the actions $a_{i,t}$ of an agent $i$, operating building $b$ and ESS $j$, are normalized as follows:

\begin{itemize}

\item \(\alpha^{\texttt{DR}}_{i,t} \in [0,1]\) specifies the fraction of building $b$’s maximum reduction capacity that is actually used at time~$t$. Concretely, the actual reduction is
\( \pred = \alpha^{\texttt{DR}}_{i,t} \overline{\mathrm{P}}^{\texttt{DR}}_b \) ensuring constraint \eqref{eq:building_bounds}.

\item \(\alpha^{\texttt{ESS}}_{i,t} \in [-1,1]\) encodes the net charging or discharging action of the building’s ESS. If \(\alpha^{\texttt{ESS}}_{i,t} = 0\), then \(\pcharge = 0\) and \(\pdischarge = 0\). If \(\alpha^{\texttt{ESS}}_{i,t} > 0\), then \(\pcharge = |\alpha^{\texttt{ESS}}_{i,t}|\,\overline{\mathrm{P}}^{\texttt{ch}}_j\) and \(\pdischarge = 0\). Finally, if \(\alpha^{\texttt{ESS}}_{i,t} < 0\), then \(\pcharge = 0\) and \(\pdischarge = |\alpha^{\texttt{ESS}}_{i,t}|\,\overline{\mathrm{P}}^{\texttt{dis}}_j\). Notice how constraints \eqref{eq:ess_charging_bounds}, \eqref{eq:ess_discharging_bounds} are always satisfied by construction.
\end{itemize}
Also, note that constraint \eqref{eq:ess_energy_bounds} is imposed by the agent's environment:  if a value of $e_{j,t}$ is below (above) limits, it is mapped back to $\underline{\mathrm{E}}_j$ ($\overline{\mathrm{E}}_j$).
Finally, we design the reward $r_{i,t}$ of an agent as
\begin{multline} \label{eq:reward_final}
    r_{i,t} = \priceflex(\pred + \pdischarge) - \\
    \pricebuy (\pcharge + \demand - \pred) - \pricered \pred - \\
    \kappa_{\texttt{DR}} \cdot \max\Big\{0,\sum_{t \in \T} \pred - \mathrm{H}_b |\T| \overline{\mathrm{P}}^{\texttt{DR}}_b \Big\}- \\
    \kappa_{\texttt{ESS}} \cdot \max \big\{0, 0.5\overline{\mathrm{E}}_j - e_{j,|\T|} \big\},
\end{multline}
which comprises the local part of the global reward (cf Eq. \eqref{eq:reward}), and two penalty terms that apply if the building exceeds its cumulative reduction limit (cf Eq. \eqref{eq:total_reduction_constraint}) or the ESS fails to meet the required state of charge by the end of the horizon (cf Eq. \eqref{eq:storage_horizon_end}).
An individual agent's policy optimization problem takes the form: 
\begin{equation}
    \max_{\pi_i} J_i(\pi_i) = \mathbb{E}\left[\sum_{t \in \T} \gamma_t r_{i,t} | \pi_i\right],
\end{equation}

In conclusion, the above design addresses constraints \eqref{eq:ess_charging_bounds}-\eqref{eq:total_reduction_constraint} of the C-MDP $\mathcal{M}$. Ensuring that the agents' actions maintain the voltage within limits (Eq \eqref{eq:volt_limits_socp}) is more challenging, given the absence of information over the distribution network, and it is addressed in the next subsection.

\subsection{Safety Layer}

The voltage constraints are addressed in two steps.
The first step is to derive a mapping from a given joint state-action profile to a voltage profile.
For this purpose, a regressor 
\begin{equation} \label{regressor}
    (\hat{v}_{n,t})_{n \in \N} = f(\bm{o}_t, \bm{a}_t; \mathbf{W}),
\end{equation}
is trained, where $\bm{o}_t, \bm{a}_t$ denote the joint agents' observations and actions and $\mathbf{W}$ is the trainable weight matrix.
Although the agents are located only at a subset of buses $\mathcal{N}_C \subseteq \mathcal{N}$, the regressor is trained in a model-free supervised learning fashion to predict voltages at all buses. This design choice leverages the full measurement data available and captures spatial interdependencies across the entire network, thereby improving prediction accuracy without relying on any knowledge of the network's topology or physical parameters.

The second step is to use the voltage regressor in a safety layer projector module, which adjusts the agents' prospective actions $\bm{a}_t$ (as little as necessary) to satisfy the voltage limits:
\begin{equation} \label{projector}
    \begin{aligned}
    & \min_{\bm{a}_t^\prime} \|\bm{a}_t^\prime - \bm{a}_t\|^2 \\
    \text{s.t.} \quad & \underline{V} \leq f(\bm{o}_t, \bm{a}_t^\prime ; \mathbf{W}) \leq \overline{V}, \quad \forall n \in \N.
    \end{aligned}
\end{equation}
To ensure an optimal solution to problem \eqref{projector}, we opt for a regressor $f(\cdot)$ that is linear in the joint actions. Note that, although we focused on voltage constraints, the framework can similarly handle other operational limits (e.g., line currents) by training additional predictors and adding corresponding constraints to the safety-layer optimization.

\subsection{The Algorithm}

We employ the Multi-Agent Deep Deterministic Policy Gradient (MADDPG) algorithm \cite{lowe2017multi} for policy optimization.
MADDPG consists of two main components for each agent: the actor and the critic. The actor is responsible for mapping the agent's local observation to an action, while the critic evaluates the expected return of taking an action in a given state, considering the current policy. Specifically, each agent \(i\) maintains a policy network \(\pi_i(o_i;\theta_i)\), parameterized by \(\theta_i\), which transforms its local observation \(o_i\) into an action \(a_i\). This \textit{actor network} is trained to maximize the expected return from the current state by applying gradients derived from the critic’s evaluation. The \textit{critic network} \(Q_i^\pi(s, a_1, \dots, a_N;\beta_i)\), parameterized by \(\beta_i\), estimates the value of the joint action in the context of the global state \(s\) and all agents’ actions \((a_1,\dots,a_N)\). Unlike the actor, the critic has access to the actions and observations of all agents, enabling it to evaluate the overall joint action value and mitigate non-stationarity during training.

During training, MADDPG stores experience tuples \((s_{t}, a_{t}, r_{t}, s_{t+1})\) in a replay buffer, enabling mini-batch updates that mitigate sequential correlations and stabilize learning. Target networks for the actor and critic provide a stable Q-value reference. The critic is updated by minimizing the gap between predicted and target Q-values, while the actor improves via policy gradients derived from the critic’s evaluation. After convergence, each agent executes its learned policy locally, coordinating load reduction and ESS actions.
%

\begin{figure}[t]
	\centering
    \includegraphics[width=0.85\columnwidth]{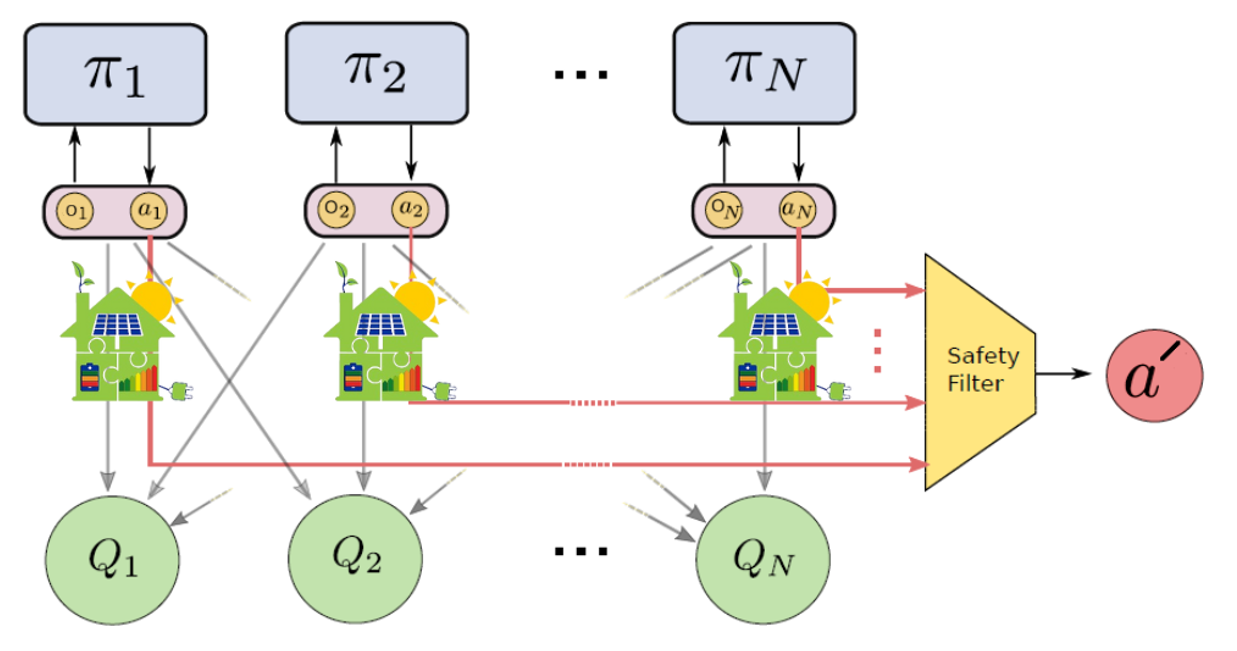}\\
	\caption{Illustration of the safety layer used in combination with the MADDPG networks.}
    \label{fig1}
\end{figure}

Fig. \ref{fig1} depicts the algorithm high-level architecture schematically, while the exact algorithm is given in Algorithm \ref{alg1}.

\begin{algorithm}[t]
    \caption{Safe MADDPG Algorithm for Flexibility Provision from Grid-Aware Buildings}\label{alg1}
    \begin{algorithmic}[1]
    \State{Initialize random weights $\beta_{i}$ and $\theta_{i}$ for each agent's critic network $Q_{i}^{\pi}(s_{t},a_{t};\beta_{i})$ and actor network $\pi_{i}(o_{i};\theta_{i})$}
    \State{Initialize weights $\hat{\beta}_{i} \gets \beta_{i}$ and $\hat{\theta}_{i} \gets \theta_{i}$ for target networks $\hat{Q}_{i}^{\pi}$ and $\hat{\pi}$}
    \State{Initialize a replay buffer $\mathcal{R} = \{\}$ to store state transitions}
    \For{episode $ = 1$ to $M$}
        \State{Randomly initialize the environment state $s_{1}$}
        \For{$t  = 1$ to $T$}
            \State{Each agent selects action $a_{i,t} = \pi_{i}(o_{i,t};\theta_{i})$}
            \State{Concatenate actions into $a_{t} = (a_{1, t},...,a_{N, t})$}
            \State{Predict resulting voltages using Eq. \eqref{regressor}}
            \State{Project actions onto the safety set, per Eq. \eqref{projector}}
            \State{Inject exploratory noise $n$ to encourage exploration}
            \State{Apply $a_{t}^\prime$, observe the reward $r_{t}$ and the next state $s_{t+1}$, including voltage changes}
            \State{Store transition $(s_{t}, a_{t}, r_{t}, s_{t+1})$ in $\mathcal{R}$ for learning}
            \State{Sample a mini-batch of $\lambda$ transitions from $\mathcal{R}$ for training}
            \State{Compute the target actions $\tilde{a}_{t}$ using the target networks for the next state}
            \State{Update critic networks by minimizing the loss between predicted and target Q-values}
            \State{Update actor networks using the policy gradient, derived from the critic's output}
            \State{Gradually update target networks using the mixing factor $\tau$}
        \EndFor
    \EndFor
    \end{algorithmic}
\end{algorithm}

\section{Case Study - Results}

\subsection{Case Study Description}

We consider the IEEE 33-bus distribution network \cite{dolatabadi2020enhanced}, which includes 33 buses and 32 lines at a nominal voltage of 12.66 kV. Five buildings are situated at buses 5, 10, 15, 20, and 25. The active and reactive load profiles have been based on \cite{wang2021multi}. Each building features a photovoltaic unit with a rated output of 150 kW and an energy storage system capable of 5 kW charging and discharging power at 90\% efficiency. The total energy capacity of the ESS is set to 25 kWh, and each building can reduce up to 50\% of its net demand, incurring a $\pricered$ of 0.01~€/kWh cost for these load reductions. 

For the flexibility price $\priceflex$ we use the day-ahead prices taken from the ENTSO-E transparency platform \cite{hirth2018entso} and expressed in~€/kWh, guiding the remuneration for reducing load or discharging the ESS. Meanwhile, the cost of buying energy $\pricebuy$ from the grid is structured as a three-level time-of-use tariff in~€/kWh, with higher rates during peak periods and lower rates during off-peak hours. 

\subsection{Voltage Regressor}

First we present the configuration of the voltage regressor and its evaluation (in isolation).
We implement a multi-output linear regression model with 33 outputs, one for each bus. The training dataset for the voltage predictor was generated by solving power flow calculations for a wide range of operational scenarios. These scenarios were created by varying active and reactive power profiles across the network.
The training process employed a 5-fold cross-validation strategy to ensure robustness and prevent overfitting. The dataset was divided into five subsets, and the model was trained and validated across different combinations of these subsets. 

\begin{figure}[t] 
    \centering 
    \includegraphics[width=0.85\columnwidth]{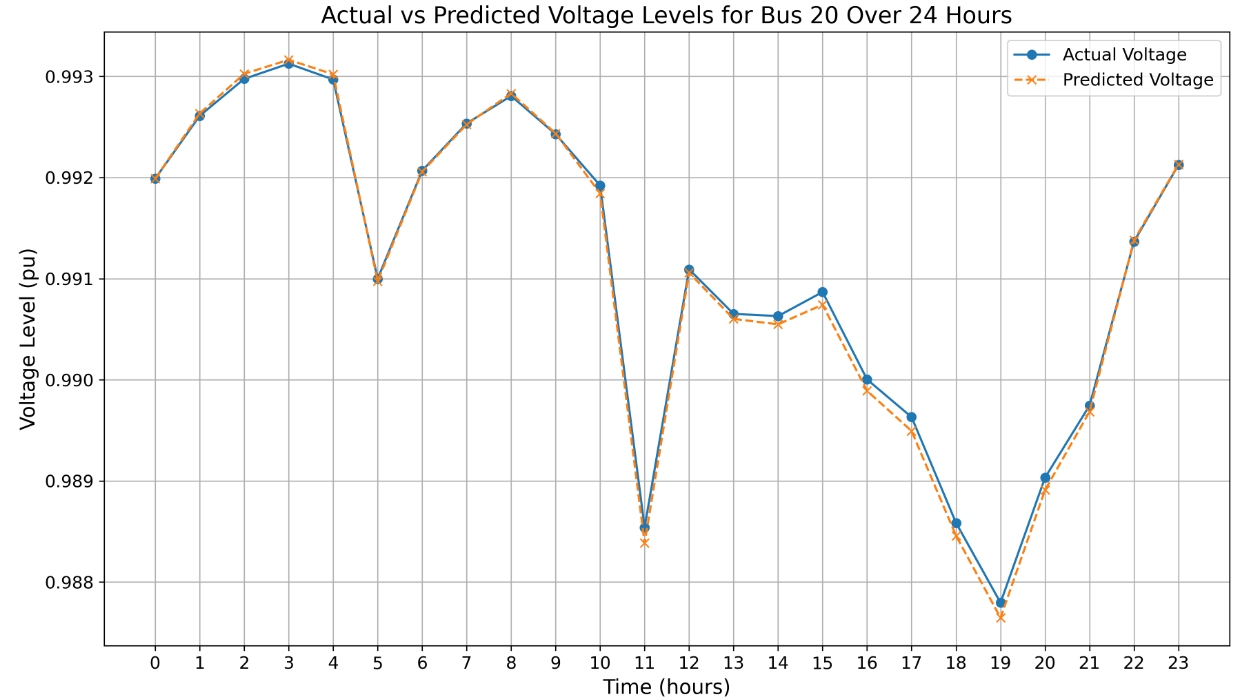} 
    \caption{Predicted and actual voltage levels for bus 5 over 24 hours.} 
    \label{fig2} 
\end{figure}

The results indicate a negligible Mean Absolute Error of approximately 0.00138 pu, which demonstrates that the predicted voltages closely match the actual measurements. Also, the 
model explains approximately 99.3\% of the variance in the voltage levels, confirming its reliability and effectiveness.
%
Fig. \ref{fig2} illustrates a comparison between the predicted and actual voltage levels at Bus 5 over a 24-hour period. The close alignment between the two curves underscores the model's precision in capturing the action-voltage relation.

\subsection{Safe MADRL}

This section presents the configuration of the proposed Safe-MADDPG algorithm and evaluates its performance while also comparing it with two benchmark algorithms: the MADDPG without the safety layer, and the multi-agent proximal policy optimization (MAPPO) algorithm. MADDPG and MAPPO handle voltage constraint violations through a penalty term subtracted from the reward function of Eq. \eqref{eq:reward_final}:

\begin{equation} \label{eq:voltage_penalty}
  \kappa_{\texttt{V}} \cdot
  \max\Bigl\{\,0,\;  \sum_{t \in \mathcal{T}}(v_{b,t} \;-\; \overline{\mathrm{V}}),\; \sum_{t \in \mathcal{T}} (\underline{\mathrm{V}} \;-\;v_{b,t})\Bigr\}.
\end{equation}

\subsubsection{Training}

The training process for the proposed algorithm and the two benchmarks was carried out for 1000 episodes utilizing two years of historical data.
The key hyperparameters for training included a discount factor \(\gamma\) of 0.99, a batch size of 32, a learning rate of \num{1e-4}, and a target network update rate \(\tau\) of 0.05. To ensure fair comparison, all three algorithms used the same architecture of a recurrent neural network for policy representation, with a hidden layer size of 64 and ReLU activation functions. Safe-MADDPG and MADDPG employed double Q-learning to improve stability, while Safe-MADDPG incorporated an additional safety layer to enforce operational constraints. The penalty coefficients used in the reward formulation are: $\kappa_{\texttt{V}}$ set to 1~(€/pu), $\kappa_{\texttt{DR}}$ set to 10~(€/kWh), and $\kappa_{\texttt{ESS}}$ set to 10~(€/kWh).

\begin{figure}[t]
    \centering
    \includegraphics[width=\columnwidth]{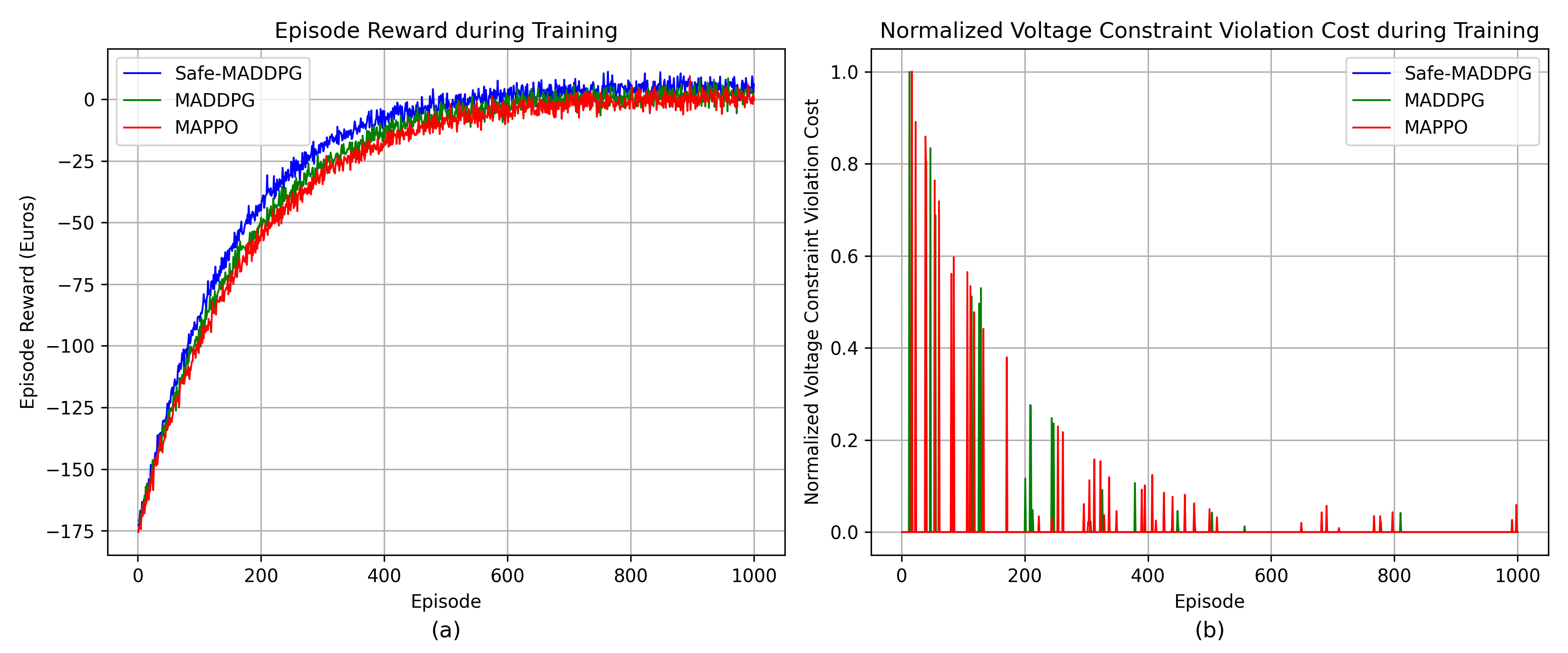}
    \caption{(a) Episode reward over training (left) and (b) normalized voltage-violation cost over training (right).}
    \label{fig3}
\end{figure}

The results in Fig.~\ref{fig3} present the training performance for the three algorithms, where subplot~(a) reports the episode rewards and subplot~(b) shows the normalized voltage-violation cost of Eq. \eqref{eq:voltage_penalty} over 1000 episodes. In subplot~(a), Safe-MADDPG consistently obtains higher episode rewards compared to MADDPG and MAPPO, stabilizing after around 600 episodes. 
Its advantage stems from the safety layer which corrects any actions leading to unsafe voltage levels, so Safe-MADDPG avoids incurring penalties for voltage violations, as in the case of MADDPG and MAPPO. By removing these violations/penalties from the learning process, Safe-MADDPG converges more efficiently and achieves closer-to-optimal policies.

Subplot~(b) highlights that, while Safe-MADDPG maintains a zero normalized violation cost throughout training, both MADDPG and MAPPO begin with relatively high violation costs but gradually reduce them after about 500 episodes. MAPPO never fully eliminates small spikes of violations, leading to a slightly higher long-term cost than MADDPG. 

\subsubsection{Testing}

The testing phase evaluates the performance of the trained policies (Safe-MADDPG, MADDPG, and MAPPO) over a one-week testing period. The results are benchmarked against the optimal-in-hindsight OPF solution, which is obtained by solving a deterministic optimization problem, counterfactually assuming that all information about uncertainties and distribution network characteristics are perfectly known. This benchmark serves as a theoretical upper bound for the objective function value. The testing considers two aspects: the resulting percentage difference from the OPF solution, and the amount of voltage constraint violations.

Table~\ref{tab:testing_results} presents the testing results. Safe-MADDPG exhibits a 12\% reduction in net benefit relative to OPF, while MADDPG and MAPPO show larger net benefit gaps of 15\% and 16\% respectively. Safe-MADDPG also maintains zero voltage violations by design, whereas MADDPG and MAPPO incur 1 and 3 violations, respectively. 

\begin{table}[t]
\centering
\caption{Testing Results}
\label{tab:testing_results}
\begin{tabular}{l|c|c}
\hline
\textbf{Algorithm} & \begin{tabular}[c]{@{}c@{}}{\% Difference}\\from OPF solution\end{tabular} & \begin{tabular}[c]{@{}c@{}}Voltage\\Violations\end{tabular} \\
\hline
Safe-MADDPG & $12\%$ & 0 \\
MADDPG      & $15\%$ & 1 \\
MAPPO       & $16\%$ & 3 \\
\hline
\end{tabular}
\end{table}



\begin{figure}[t] 
    \centering 
    \includegraphics[width=0.85\columnwidth]{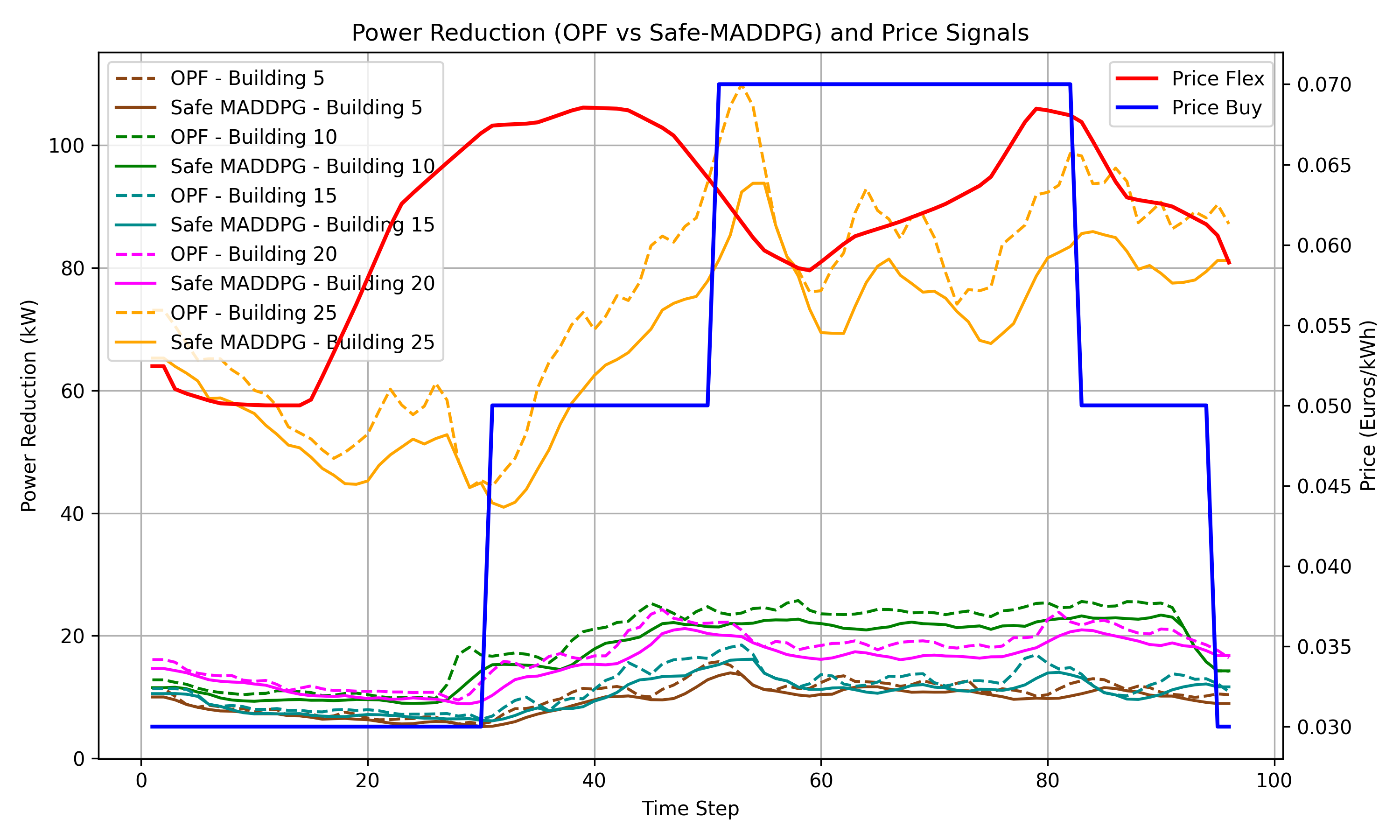} 
    \caption{Comparison of power reduction per building between the OPF solution and the Safe-MADDPG policy for a single day, alongside flexibility price and buy prices.} 
    \label{fig4} 
\end{figure}

Additionally, a comparative analysis is performed to showcase the behavior of the Safe-MADDPG policy against the OPF solution in terms of power reduction and ESS energy.
Fig. \ref{fig4} illustrates the power reduction achieved for each building and the corresponding DR flexibility price as well as the time-of-use tariff during the testing period. The OPF solution and the Safe-MADDPG policy follow a similar pattern, demonstrating a more conservative nature of the Safe-MADDPG policy in decision-making compared to OPF: while the OPF solution pushes for maximum power reduction to the limits posed by the constraint (\ref{eq:building_bounds}) to optimize flexibility costs, the Safe-MADDPG policy adopts a more cautious approach, slightly limiting power reduction.

\begin{figure}[t] 
    \centering 
    \includegraphics[width=0.85\columnwidth]{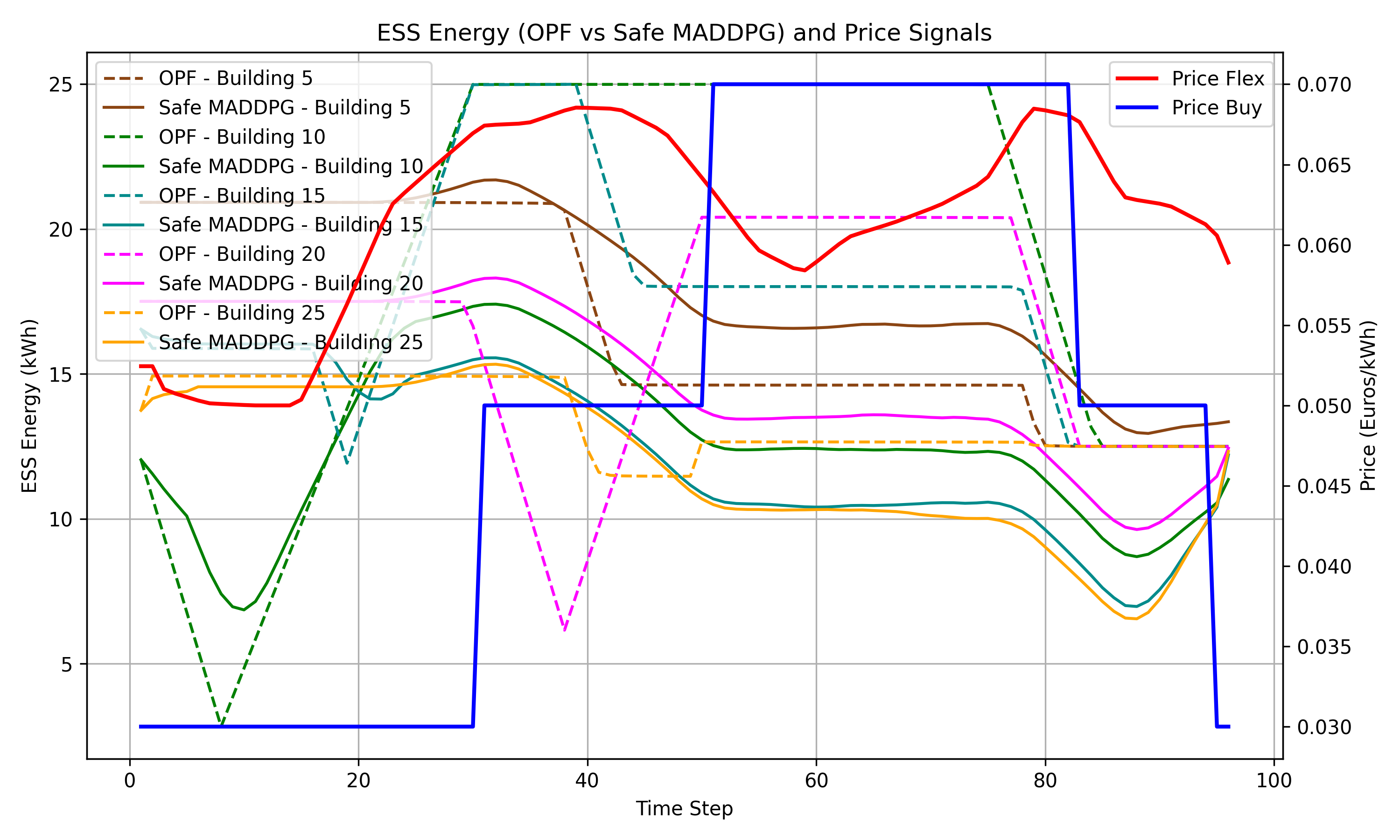} 
    \caption{Comparison of ESS energy levels per building between the OPF solution and the Safe-MADDPG policy for a single day, alongside flexibility and buy prices.} 
    \label{fig5} 
\end{figure}

Fig. \ref{fig5} depicts the ESS energy levels and the corresponding DR flexibility price as well as the time-of-use tariff during the testing period. 
The OPF solution and the Safe-MADDPG policy exhibit similar trends, with the Safe-MADDPG policy demonstrating a more conservative approach in charging and discharging decisions. While the OPF solution fully utilizes the ESS capacity to optimize operational costs, the Safe-MADDPG policy prioritizes safety by maintaining energy levels within safer operational bounds.

The comparisons above indicate that the Safe-MADDPG policy is able to track the pattern of the optimal-in-hindsight OPF solution while (given the uncertainties) it is holding back on aggressive actions to avoid violating the intertemporal (delayed) constraints relating to end-of-horizon storage and cumulative demand-reduction limits. 


\section{Conclusions}
This paper presented a safe MARL policy for making bottom-up DER management decisions related to flexibility services.
A safety layer was introduced to the policy, ensuring that voltage limits are respected.
Towards overcoming the limiting assumption of available network information, the safety layer was based on a model-free voltage predictor relying solely on local voltage measurements.
The simulation results indicated that the proposed policy enables self-organized DER communities to provide distribution-network-safe flexibility services without requiring in-the-loop DSO engagement, whilst outperforming state-of-the-art benchmarks. Future work will investigate voltage prediction under scenarios with partial observability, reflecting realistic conditions with incomplete measurement availability, as well as extension to multi‑period bidding in ancillary‑service markets.

\section*{Acknowledgment}
This work was supported by European Union’s funded projects DIGITISE (grant number 101160671), ARV (grant number 101036723) and ELEXIA (grant number 101075656).

\bibliographystyle{elsarticle-num} 
\bibliography{references}

\begin{thebibliography}{10}
\expandafter\ifx\csname url\endcsname\relax
  \def\url#1{\texttt{#1}}\fi
\expandafter\ifx\csname urlprefix\endcsname\relax\def\urlprefix{URL }\fi
\expandafter\ifx\csname href\endcsname\relax
  \def\href#1#2{#2} \def\path#1{#1}\fi

\bibitem{ceer2020ceer}
C.~D. S.~W. Group, et~al., {CEER} paper on {DSO} procedures of procurement of flexibility, Council of European Energy Regulators (CEER), Brussels, Report C19-DS-55--05 (2020).

\bibitem{molzahn2017survey}
D.~K. Molzahn, F.~D{\"o}rfler, H.~Sandberg, S.~H. Low, S.~Chakrabarti, R.~Baldick, J.~Lavaei, A survey of distributed optimization and control algorithms for electric power systems, IEEE Transactions on Smart Grid 8~(6) (2017) 2941--2962.

\bibitem{tsaousoglou2023integrating}
G.~Tsaousoglou, R.~Junker, M.~Banaei, S.~S. Tohidi, H.~Madsen, Integrating distributed flexibility into {TSO-DSO} coordinated electricity markets, IEEE Transactions on Energy Markets, Policy and Regulation (2023).

\bibitem{zhang2022multi}
B.~Zhang, W.~Hu, A.~M. Ghias, X.~Xu, Z.~Chen, Multi-agent deep reinforcement learning-based coordination control for grid-aware multi-buildings, Applied Energy 328 (2022) 120215.

\bibitem{charbonnier2022scalable}
F.~Charbonnier, T.~Morstyn, M.~D. McCulloch, Scalable multi-agent reinforcement learning for distributed control of residential energy flexibility, Applied Energy 314 (2022) 118825.

\bibitem{ye2022multi}
Y.~Ye, D.~Papadaskalopoulos, Q.~Yuan, Y.~Tang, G.~Strbac, Multi-agent deep reinforcement learning for coordinated energy trading and flexibility services provision in local electricity markets, IEEE Transactions on Smart Grid (2022).

\bibitem{elsayed2021safe}
I.~ElSayed-Aly, S.~Bharadwaj, C.~Amato, R.~Ehlers, U.~Topcu, L.~Feng, Safe multi-agent reinforcement learning via shielding, arXiv preprint arXiv:2101.11196 (2021).

\bibitem{gu2021multi}
S.~Gu, J.~G. Kuba, M.~Wen, R.~Chen, Z.~Wang, Z.~Tian, J.~Wang, A.~Knoll, Y.~Yang, Multi-agent constrained policy optimisation, arXiv preprint arXiv:2110.02793 (2021).

\bibitem{lu2021decentralized}
S.~Lu, K.~Zhang, T.~Chen, T.~Ba{\c{s}}ar, L.~Horesh, Decentralized policy gradient descent ascent for safe multi-agent reinforcement learning, in: Proceedings of the AAAI Conference on Artificial Intelligence, Vol.~35, 2021, pp. 8767--8775.

\bibitem{sheebaelhamd2021safe}
Z.~Sheebaelhamd, K.~Zisis, A.~Nisioti, D.~Gkouletsos, D.~Pavllo, J.~Kohler, Safe deep reinforcement learning for multi-agent systems with continuous action spaces, arXiv preprint arXiv:2108.03952 (2021).

\bibitem{wang2023secure}
Y.~Wang, D.~Qiu, M.~Sun, G.~Strbac, Z.~Gao, Secure energy management of multi-energy microgrid: A physical-informed safe reinforcement learning approach, Applied Energy 335 (2023) 120759.

\bibitem{gao2022model}
Y.~Gao, N.~Yu, Model-augmented safe reinforcement learning for volt-var control in power distribution networks, Applied Energy 313 (2022) 118762.

\bibitem{zhang2024networked}
J.~Zhang, L.~Sang, Y.~Xu, H.~Sun, Networked multiagent-based safe reinforcement learning for low-carbon demand management in distribution networks, IEEE Transactions on Sustainable Energy (2024).

\bibitem{chen2022physics}
P.~Chen, S.~Liu, X.~Wang, I.~Kamwa, Physics-shielded multi-agent deep reinforcement learning for safe active voltage control with photovoltaic/battery energy storage systems, IEEE Transactions on Smart Grid (2022).

\bibitem{guo2023safe}
G.~Guo, M.~Zhang, Y.~Gong, Q.~Xu, Safe multi-agent deep reinforcement learning for real-time decentralized control of inverter based renewable energy resources considering communication delay, Applied Energy 349 (2023) 121648.

\bibitem{zhang2020multi}
Q.~Zhang, K.~Dehghanpour, Z.~Wang, F.~Qiu, D.~Zhao, Multi-agent safe policy learning for power management of networked microgrids, IEEE Transactions on Smart Grid 12~(2) (2020) 1048--1062.

\bibitem{lowe2017multi}
R.~Lowe, Y.~I. Wu, A.~Tamar, J.~Harb, O.~Pieter~Abbeel, I.~Mordatch, Multi-agent actor-critic for mixed cooperative-competitive environments, Advances in neural information processing systems 30 (2017).

\bibitem{dolatabadi2020enhanced}
S.~H. Dolatabadi, M.~Ghorbanian, P.~Siano, N.~D. Hatziargyriou, An enhanced {IEEE} 33 bus benchmark test system for distribution system studies, IEEE Transactions on Power Systems 36~(3) (2020) 2565--2572.

\bibitem{wang2021multi}
J.~Wang, W.~Xu, Y.~Gu, W.~Song, T.~C. Green, Multi-agent reinforcement learning for active voltage control on power distribution networks, Advances in Neural Information Processing Systems 34 (2021) 3271--3284.

\bibitem{hirth2018entso}
L.~Hirth, J.~M{\"u}hlenpfordt, M.~Bulkeley, The {ENTSO-E} transparency platform--a review of europe’s most ambitious electricity data platform, Applied energy 225 (2018) 1054--1067.

\end{thebibliography}

\end{document}